\documentclass[aps,prl,10pt,amsmath,twocolumn,superscriptaddress]{revtex4-1}
\usepackage{graphicx}
\usepackage{subfigure}

\begin{document}

\title{Remote Dipolar Interactions for Objective Density Calibration and Flow Control of Excitonic Fluids}
\author{Kobi Cohen}
\author{Ronen Rapaport}
\affiliation{Racah Institute of Physics, The Hebrew University of Jerusalem, Jerusalem 91904, Israel}
\author{Paulo V. Santos}
\affiliation{Paul-Drude-Institut f\"{u}r Festk\"{o}rperelektronik, Hausvogteiplatz 5-7, 10117 Berlin, Germany}

\begin{abstract}
In this paper we suggest a method to observe \emph{remote} interactions of spatially separated dipolar quantum fluids, and in particular of dipolar excitons in GaAs bilayer based devices. The method utilizes the static electric dipole moment of trapped dipolar fluids to induce a local potential change on spatially separated test dipoles. We show that such an interaction can be used for a model-independent, objective fluid density measurements, an outstanding problem in this field of research, as well as for inter-fluid exciton flow control and trapping. For a demonstration of the effects on realistic devices, we use a full two-dimensional hydrodynamical model.
\end{abstract}

\maketitle

Dipolar quantum fluids are very intriguing physical entities that are intensively sought for in various material systems, such as ultra-cold polar molecules \cite{SantosZoller00} and excitons in semiconductor bilayer structures \cite{Snoke02, Eisenstein04}. The interest in such fluids arises from the basic unresolved questions regarding their possible rich thermodynamic phases, which are predicted to emerge from the interplay between their quantum statistics and the particular form of the dipole-dipole interaction \cite{BuchlerZoller07, Lozovik07, Laikhtman09}. While the local interactions between nearby dipoles is expected to determine the physical state of the fluid, it should be interesting to observe the effect of interactions between spatially remote fluids. This concept of remote interactions can also be utilized, as we will suggest here, in a couple of aspects: it can be used to determine in an objective, model-independent way the density of a dipolar fluid, a tricky "boot-strap" problem, especially for excitons in bilayer systems. It can also be utilized as a method to control the flow of one fluid by other, remotely located fluids.

\begin{figure}[h]
\subfigure{\includegraphics[width=0.35\textwidth]{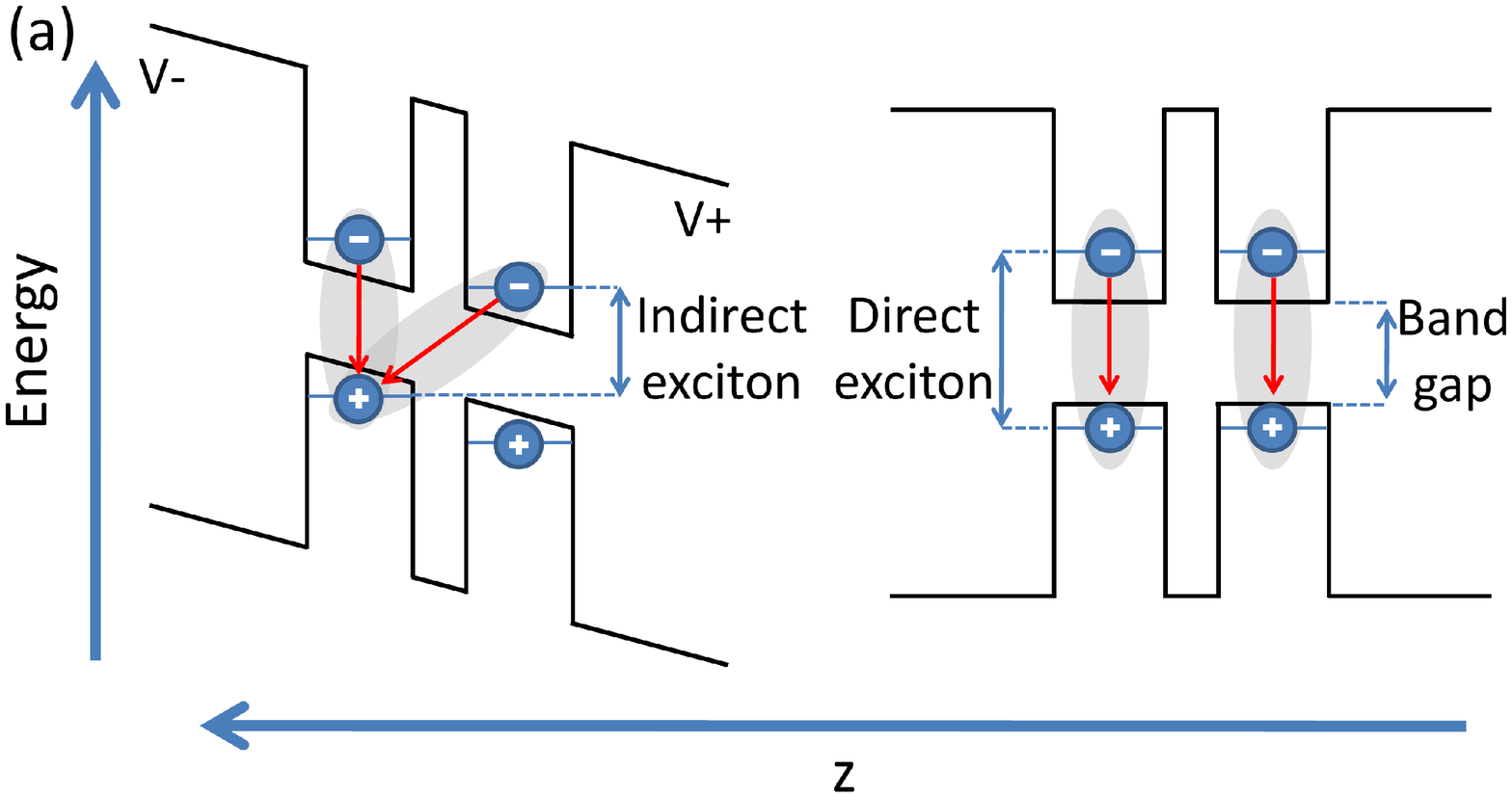}}
\subfigure{\includegraphics[width=0.35\textwidth]{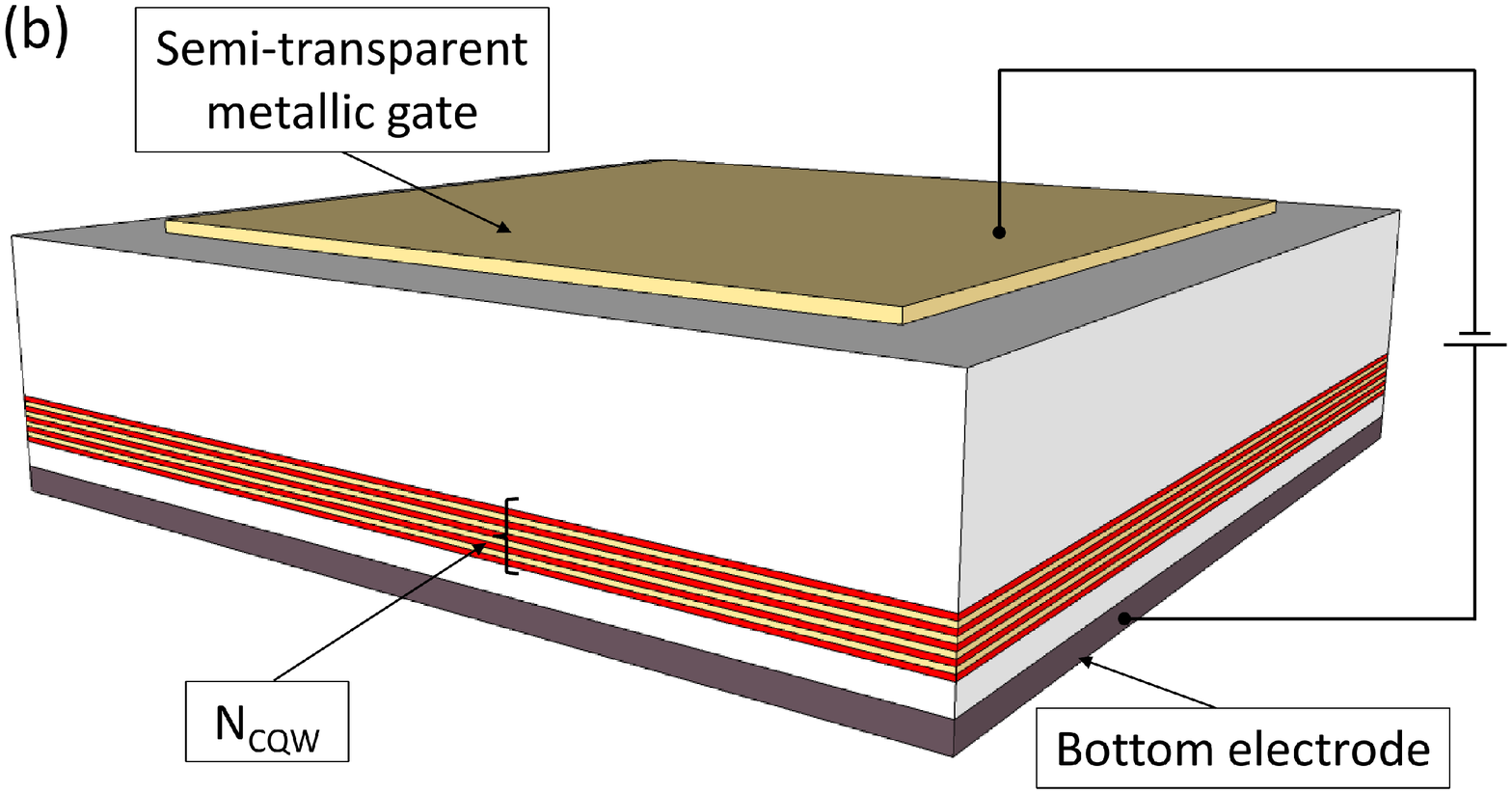}}\\
\caption{(a) A Schematic band diagram for a GaAs/AlGaAs CQW structure without (right) and with (left) external bias, showing the formation of indirect (dipolar) excitons. (b) A schematic sketch of the proposed sample structure, containing a stack of CQWs (their number is marked by $N_{CQW}$), a homogeneous back electrode and top semitransparent electric gate. The specific top gate geometries for the various implementations discussed in the text are given in Fig. \ref{fig:device_sketch}.}
\label{fig:dipolar_exciton}
\end{figure}

As mentioned above, dipolar fluids can be realized for excitons in bilayer structures. An exciton consists of a pair of an electron and a hole inside a semiconductor, which are bound together by their mutual coulomb attraction, forming a boson-like quasi particle. In electrically gated GaAs coupled quantum wells (CQW), also known as bilayer structures, an externally applied electric field, using semi-transparent electrical gates, results in the creation of two-dimensional dipolar excitons. These light mass excitonic quasi-particles have their electron and hole constituents separated into two layers (Fig. \ref{fig:dipolar_exciton}(a)) and thus have an increased life time, an elevated quantum degeneracy temperature (compared to atoms or molecules for example), and all carry a dipole moment oriented perpendicular to the QW plane. The two dimensional dipolar excitons $(X_d)$ interact with each other via the repulsive dipole-dipole forces, and therefore exciton based systems are model systems for exploring a low dimensional interacting fluid hydrodynamics at quantum degeneracy conditions. The $X_d$ can also be transformed back into light signals in an almost perfect on/off controlled fashion by switching the external gates electric fields, thus can potentially be used for unique functional optoelectronic devices.
Various methods can also be used to effectively transport and manipulate $X_d$ flows. Indeed, in the past few years several groups introduced relatively efficient ways of controlling $X_d$ fluxes, by using electric field gradients \cite{Holleitner06}, surface acoustic waves \cite{Rudolph07} and modulated gate voltages \cite{High08,ButovSwitch09,Kuznetsova10}. Transport of excitons has also been measured in other bilayer systems \cite{Eisenstein00,Shayegan04,Snoke06}. However, all of the above methods are based on external manipulations and not on the internal state of the device. This may limit future functionality of devices based on exciton fluids \cite{High08,Kuznetsova10}.

Another crucial experimental difficulty, which is important in the aspect of the fundamental understanding of the $X_d$ fluid thermodynamics, is to independently and accurately estimate the density of $X_d$ created inside a sample. Such an estimate is vital for correctly identifying and gaining physical insight into phenomena such as phase transitions, quantum-degeneracy and particle correlations. This task however, turns out to be not straight-forward. The current method for estimating $X_d$ densities is to deduce it from measurements of the energy shift ("blue shift") of recombining $X_d$'s photoluminescence (PL). This energy shift strongly depends on the local density correlations around the recombining excitons, which in turn depends on the thermodynamic state of the fluid \cite{Zimmermann08,Laikhtman09}. Therefore this method relies on a knowledge of the fluid state, which makes this density calibration method a strongly model-dependant, "boot-strap" process. Finally, we note that while dipole-dipole interactions have been observed to affect the properties and the dynamics of a dipolar fluid locally, the remote effect of a collection of such dipoles on a system spatially separated from it has not been observed as far as we know. This concept alone is worthwhile pursuing.

Here, we propose that remote interactions between spatially separated excitons can be observed in both spectroscopy and hydrodynamics experiments. Such an observation can also be used for resolving practical current challenges as will be suggested in this paper: in the first part of the paper we describe a scheme that utilizes remote interactions for an objective, model-independent dipolar fluid density measurements which does not depend on local density correlations, and in the second part we describe a method to manipulate dipolar fluids using their remote interactions. We note that while here we present the implementation of the concept for dipolar exciton fluids in electrostatic traps, the concept developed here is quite general and can be implemented in other types of trapping devices and also for different types of dipolar fluids.

\begin{figure}[h]
\subfigure{\includegraphics[width=0.4\textwidth]{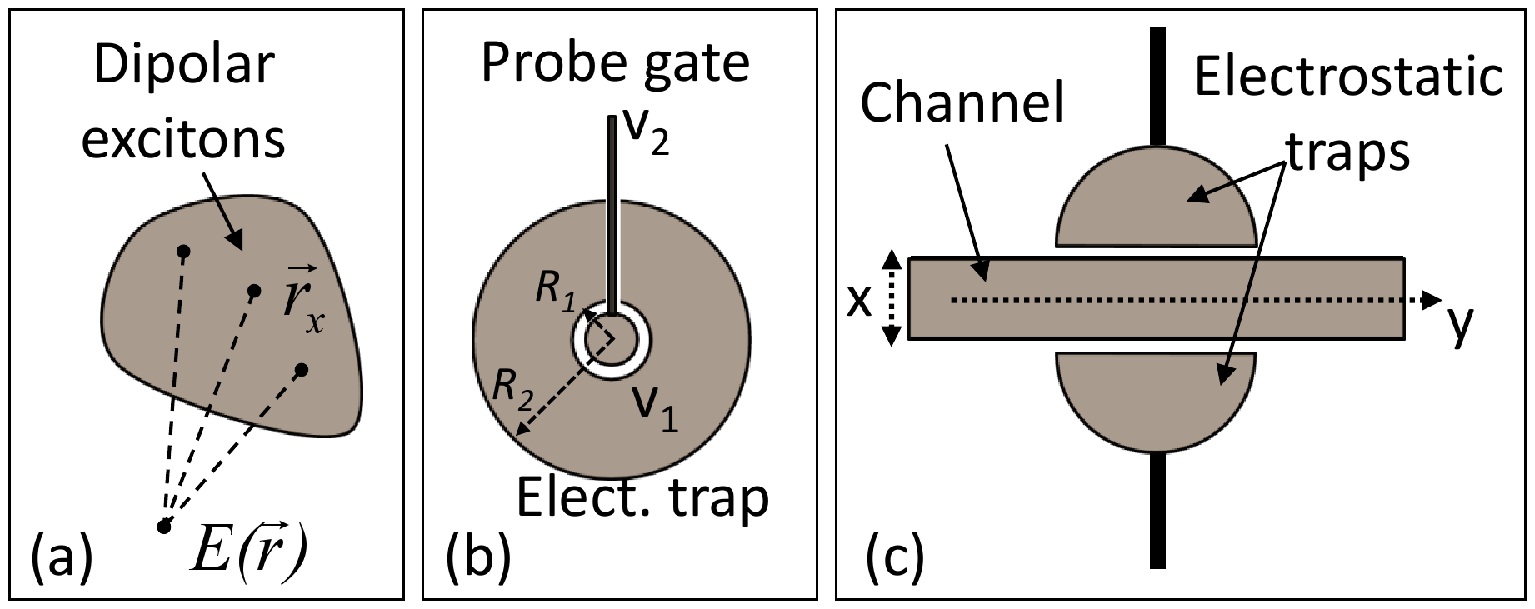}}\\
\vspace{-0.3cm}
\subfigure{\includegraphics[width=0.4\textwidth]{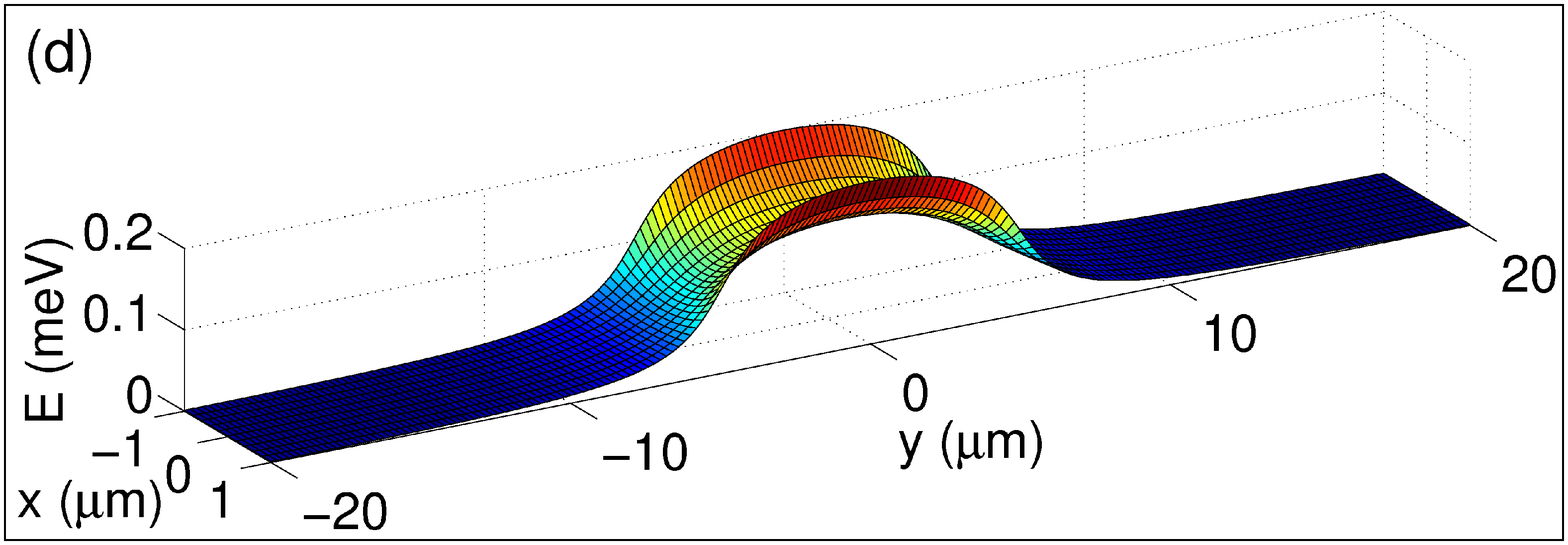}}
\caption{(color online). (a) The remote interaction energy $E$ at a point $\vec{r}$ is calculated by integrating the contribution from $X_d$ contained inside a specified region. (b) A proposed exciton density measurement device. The detected emission energy of $X_d$ trapped inside the inner probe gate is shifted due to the interaction with $X_d$ in the outer trap. $V_1$ and $V_2$ are the applied voltages on the probe and outer trap respectively. (c) A sketch of a flow device containing a narrow channel, surrounded by two half-circular electrostatic traps. (d) A numerical calculation of the energy profile along a $2\mu m$-wide channel of the device in (c), formed by $X_d$ in the two surrounding $10\mu m$ diameter electrostatic traps, which are located $1\mu m$ away from the channel edges. Here, $d=16nm$, $N_{CQW}=4$ and $n=10^{11} cm^{-2}$.}
\label{fig:device_sketch}
\end{figure}

As a starting point for our discussion, one has to consider a trap region which will confine the $X_d$ in the QW plane, and oppose their spreading due to diffusion and repulsive interaction. Various ways for trapping dipolar excitons have been realized in the past few years \cite{Zimmerman97Trap,Butov06Trap1,Butov06Trap2}. The approach we took in the current work utilizes electrostatic traps, which have a flat potential profile and a sharp boundary \cite{Rapaport05,Rapaport06}, leading to a flat average density distribution of excitons in the trap \textit{on a macroscopic scale}, denoted here by $n$ (note that local, microscopic density inhomogeneities are always expected due to dipole-dipole interactions \cite{Laikhtman09}). These electrostatic gates can also define narrow channels, in which excitons can propagate only in one direction \cite{Holleitner07}. In the following discussion, we consider a sample containing a vertical stack of separated CQWs (Fig. \ref{fig:dipolar_exciton}(b)). The number of CQW is denoted by $N_{CQW}$. As will be shown hereafter, increasing $N_{CQW}$ will increase the effect of the remote interactions. We let $d$ denote the dipole length, i.e. the spatial separation of the electron and hole wave functions in each CQW along the $z$ axis, $\vec{r}$ is a radius vector in the QW plane denoting the position of a test dipole and $\vec{r}_x$ is an arbitrary point inside the trapping region. In terms of the interaction energy $U_d(\vec{r}-\vec{r}_x)$ between two dipoles separated by a distance $|\vec{r}-\vec{r}_x|$, the total interaction energy for a test dipole at a point $\vec{r}$ is given by \cite{Laikhtman09}:
\begin{equation}\label{eq:total_interaction_with_correlation}
    E_d(\vec{r})=N_{CQW}\cdot n\iint\limits_{trap}{U_d(\vec{r}-\vec{r}_x)g_d(\vec{r}-\vec{r}_x)\text{d}^2r_x},
\end{equation}

where $g_d(\vec{r}-\vec{r}_x)$ is the pair correlation function and $ng_d(\vec{r}-\vec{r}_x)\text{d} ^2r_x$ is the average number of $X_d$'s within an area $\text{d} ^2r_x$ at a distance $|\vec{r}-\vec{r}_x|$ from the test dipole. It is clear that $E_d(\vec{r})$ for any $\vec{r}$ inside the trap strongly depends on the shape of the correlation function $g_d$. Therefore, extracting $n$ from a local measurement of $E_d(\vec{r})$ requires an independent knowledge of $g_d$. This is the essence of the objective density calibration problem. However, $g_d(\vec{r}-\vec{r}_x)|_{|\vec{r}-\vec{r}_x|\rightarrow\infty}=1$, as at large distances all local correlations disappear. Hence, for a test dipole remotely located from the rest of the trapped $X_d$ fluid, as shown schematically in Fig. \ref{fig:device_sketch}(a), Eq. \ref{eq:total_interaction_with_correlation} reduces to:

\begin{equation}\label{eq:total_interaction}
    E_d(\vec{r})=N_{CQW}\cdot n\iint\limits_{trap}{U_d(\vec{r}-\vec{r}_x)\text{d} ^2r_x}.
\end{equation}

Under such conditions, the dependence of $E_d(\vec{r})$ on $g_d$ vanishes and no knowledge of the form of $g_d$ is required to extract $n$ from a measurement of $E_d(\vec{r})$.

To check this concept of dipolar fluid density measurements in realistic devices, we consider the geometry shown in Fig. \ref{fig:device_sketch}(b). By applying two different voltages on the circular probe gate and the electrostatic ring trap surrounding it, two potential wells for the dipolar excitons in the QW plane are formed, one under each gate. Due to the radial gap between the two separated electrodes, these potential wells are spatially separated by a narrow barrier. Furthermore, as these potential wells have different depths (due to different applied gate voltages $V_1$ and $V_2$), their corresponding $X_d$ PL can be spectrally separated \cite{Rapaport05}. A straight-forward calculation for $E_d$ in this case using Eq. \eqref{eq:total_interaction} yields the result:

\begin{equation}\label{eq:probe_gate}
    E_d(\vec{r}=0)=N_{CQW}\cdot 2\pi\frac{ne^2d^2}{\epsilon}\left(\frac{1}{R_1}-\frac{1}{R_2}\right)
\end{equation}

where $R_1 \text{ and } R_2$ are the inner and outer radii of the electrostatic ring trap respectively, and we have used $U_d=e^2d^2/(\epsilon |\vec{r}-\vec{r}_x|^3)$, where $\epsilon$ is the QW dielectric constant. This approximation for $U_d$ is valid if $|\vec{r}-\vec{r}_x|\gg \!d$ for every $\vec{r}_x$, which is the case we are considering here. Increasing $N_{CQW}$ linearly increases the effective density at each point $\vec{r}_x$ and thus the interaction strength \footnote{The linear dependence of Eq. \ref{eq:total_interaction} on $N_{CQW}$ is correct as long as the vertical size of the CQW stack is much smaller than the  minimal in-plane distance $|\vec{r}_x-\vec{r}|$, a condition which can easily be met}. Substituting typical values of $N_{CQW}=5$, $n=10^{11}cm^{-2}$, $d=20nm$, $R_1=4\mu m$, $R_2=50\mu m$ into Eq. \eqref{eq:probe_gate} gives $E_d\approx0.3meV$. This $E_d$ term can be detected in the PL line blue shift of the probe $X_d$, and can be used to deduce the $X_d$ density $n$ inside the outer trap, as all other parameters are known. Since the microscopic density correlation scale $r_c$ is typically smaller or of the order of the inter-particle distance $\sim n^{-1/2}$, both $R_1$ and $R_2$ are much larger than $r_c$. As a result, this is a model-independent density measurement, as it does not assume anything on the local state of the $X_d$ fluid in the outer ring. The only assumption is that $n$ is constant on the macroscopic scale of the trap area, which was already experimentally verified \cite{Rapaport05,Rapaport06}.

While in the first part we showed that the effect of remote dipolar interactions can be observed spectroscopically, we now suggest n hydrodynamic effect through the concept of flow control using remote interactions. We consider a flow device schematically shown in Fig. \ref{fig:device_sketch}(c). Here, a $2\mu m$-wide channel gate is surrounded by two half circular electrostatic traps with a $10\mu m$ diameter each. The traps are loaded with a steady state density $n$. Applying Eq. \eqref{eq:total_interaction} to all points inside the channel, yields an interaction energy profile, shown in Fig. \ref{fig:device_sketch}(d). For typical experimental parameters $d=16nm, n=10^{11}cm^{-2}, N_{CQW}=4$ and $\epsilon=\epsilon_{GaAs}=13$, the peak has a value of $\approx 0.15meV$. We denote this maximal height of the energy "bump" by $E_b$.

\begin{figure}[h]
\qquad
\subfigure{\includegraphics[width=0.2\textwidth]{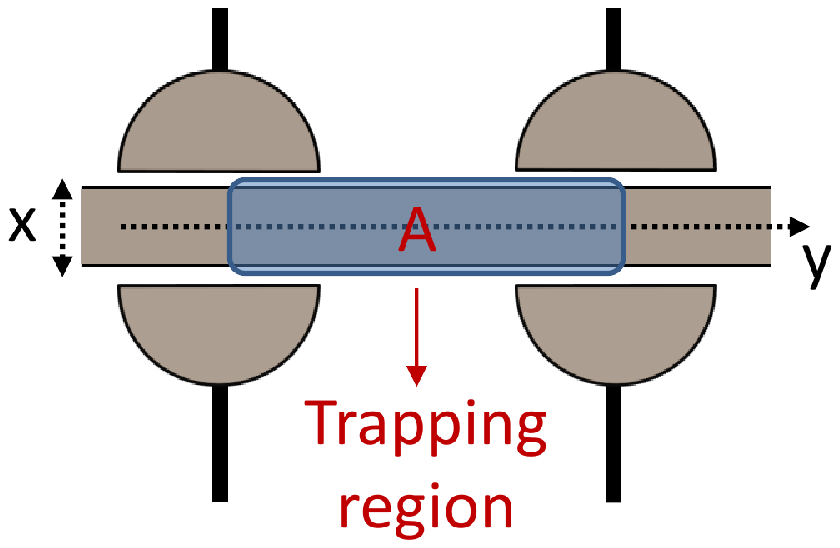}}\\
\vspace{-0.3cm}
\subfigure{\includegraphics[width=0.4\textwidth]{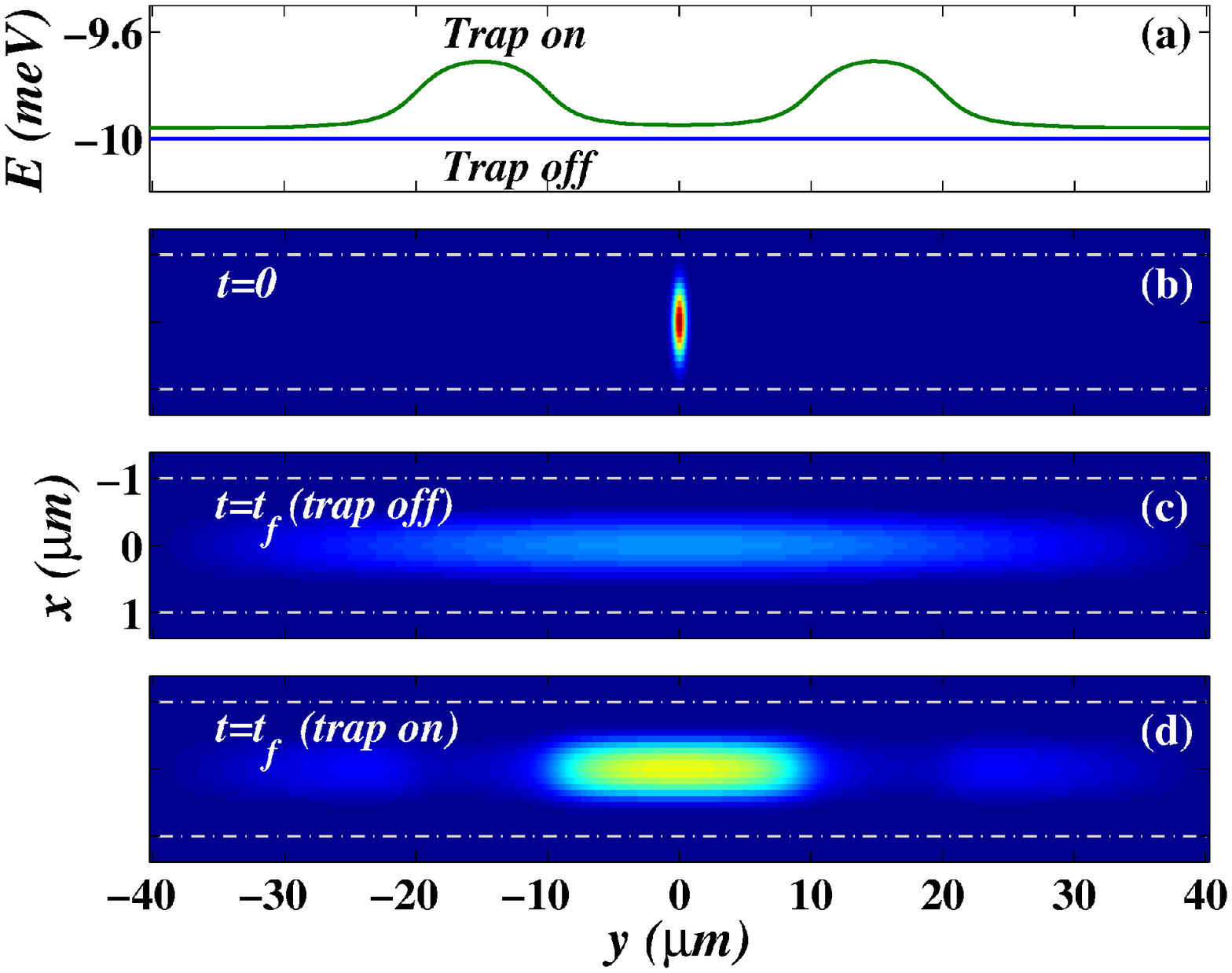}}
\caption{(color online). Upper panel: a proposed trapping scheme that consists of a narrow channel along the y direction, and 4 adjacent control traps. The effective trapping region, $A$, is marked by the shaded rectangle. Lower panels: (a) The energy profile along the channel with trapping on (green line), and off (blue line). (b) Initial $X_d$ distribution in the channel at $t=0$. The dashed lines mark the channel edges. The resulting $X_d$ distribution for $t_f=2\mu s$ is shown (color scale $\times50$) for (c) trapping "off" (empty control traps) and (d) trapping "on" (occupied control traps), showing a strong suppression of the $X_d$ spreading along the channel.}
\label{fig:trapping_2D}
\end{figure}

We now use an extension of the device shown in Fig. \ref{fig:device_sketch}(c) to demonstrate efficient trapping of $X_d$ in a channel. This extension is described in Fig. \ref{fig:trapping_2D}, where the trapping scheme consists of a narrow channel of $2\mu m$ width, and four adjacent half-circular electrostatic control traps of diameter $10\mu m$, located $1\mu m$ away from the channel edges. The control traps are again loaded with a steady state $X_d$ density (by, e.g., CW excitation) which can be tuned to control the energy profile along the channel, denoted by $E(\vec{r})$. The traps' location thus defines a trapping region along the cannel (marked by $A$), with a length of $\approx 30\mu m$.

To study the dynamics of the proposed device, we use a numerical integration of the exciton hydrodynamics equation\cite{Ivanov02,Rapaport07}:
\begin{equation}\label{eq:n}
    \frac{\partial n}{\partial t}+\nabla\left(\vec{J}_D+\vec{J}_d+\vec{J}_{ext}\right)+\frac{n}{\tau}=0
\end{equation}
Here, $n$ and $\tau$ denote the (time and spatially dependent) $X_d$ density and lifetime, respectively. The terms appearing inside the brackets are the diffusion induced current $\vec{J}_D=-D\nabla n$,
the dipolar repulsion induced current $\vec{J}_d=-n\mu\alpha\nabla n$ and a current due to external forces $\vec{J}_{ext}=n\mu \nabla E$, where $\mu$ is the exciton mobility and $\alpha$ is a material and structure parameter (see ref. \cite{Rapaport07}). We also assumed a uniform temperature of the excitons throughout the sample.
The principle of remote interactions is demonstrated by trapping freely expanding $X_d$ inside the one dimensional channel, as described in Fig. \ref{fig:trapping_2D}. The double-bump energy profile along the channel is created by loading the control traps with $X_d$ (Fig. \ref{fig:trapping_2D}(a)). An initial gaussian distribution spot $n_i(\vec{r})=n_i \exp(-4r^2)$ of $X_d$ is created inside the channel at $t=0$ (Fig. \ref{fig:trapping_2D}(b)), and the system evolves in time according to Eq. \eqref{eq:n} to $t_f=2\mu s$. The resulting $X_d$ density profiles (Fig. \ref{fig:trapping_2D}(c,d)) show that the $X_d$ are much better confined along the channel when the control traps are loaded (trapping "on").
\begin{figure}[h]
\includegraphics[width=0.45\textwidth]{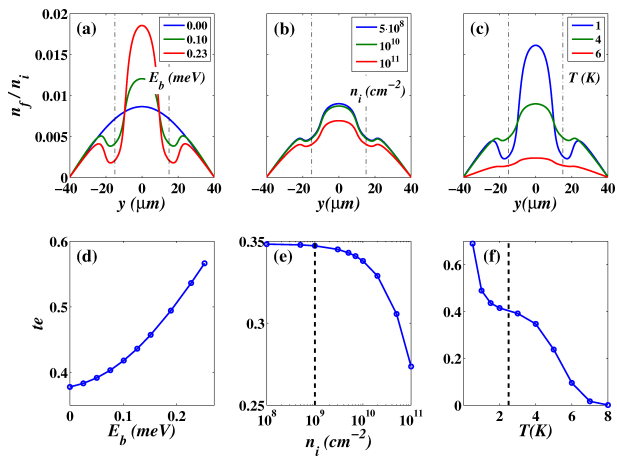}
\caption{(color online). The normalized $X_d$ density, $n_f/n_i$, at $t_f=2\mu s$ for changing (a) bump height $E_b$, (b) initial density $n_i$ and (c) temperature $T$. The trapping efficiency \textit{te} as a function of the same parameters as in (a-c) is shown in (d-f). The dashed lines in (e) and (f) marks the density and temperature, respectively, where $\vec{J}_{ext}\approx\vec{J}_d+\vec{J}_D$.}
\label{fig:trap_parameter}
\end{figure}

To characterize and quantify the performance of such a flow control device, we analyzed the same device configuration described in Fig. \ref{fig:trapping_2D}, but under various conditions. Figure \ref{fig:trap_parameter}(a-c) shows the resulting $X_d$'s density profile, $n_f(\vec{r})$, along the channel, normalized to the maximal initial $X_d$ density $n_i$, and its dependence on $E_b$, $n_i$ and the temperature $T$. One can see that $n_f$ increases with the bump height $E_b$, and decreases with the initial $X_d$'s density $n_i$ and temperature $T$ \footnote{For varying $E_b$, we used $n_i=10^9 cm^{-2}$ and $T=1.5K$. In the case of varying the initial densities, we used $E_b=0.125meV$ and $T=4K$. For varying temperatures, we set $n_i=10^9 cm^{-2}$ and $E_b=0.125meV$.}.
An objective measure for quantifying the performance of the device is the effective trapping efficiency (\textit{te}), defined as
\begin{equation}\label{eq:te}
    te=\frac{\int\limits_{A}n_f(\vec{r})\text{d}^2\vec{r}}{\int\limits_{A}n_i(\vec{r})\text{d}^2\vec{r}}
    \cdot\exp(t_f/\tau),
\end{equation}
where the integral is over the trapping area $A$ (as marked in Fig. \ref{fig:trapping_2D}) and the factor $\exp(t_f/\tau)$ compensates for the recombination loss of particles due to their finite lifetime $\tau$, thus making the above definition of \textit{te} a time-independent one (excluding other loss mechanisms such as boundary effects).
The results show that there is an increase in \textit{te} when $E_b$ is raised (Fig. \ref{fig:trap_parameter}(d)), and a decrease in \textit{te} with increasing $n_i$ ((Fig. \ref{fig:trap_parameter}(e)) and increasing temperature (Fig. \ref{fig:trap_parameter}(f)).

We can qualitatively understand these results by means of comparing the different $X_d$ currents.
For an energy bump $E_b$ with a typical width $W$, the induced current is estimated by:
$\vec{J}_{ext}=n\mu \nabla E \approx n_i\mu E_b/W$,
while for an initial $X_d$ density profile $n_i(\vec{r})$ with typical width $L$, the dipolar repulsion current will be:
$\vec{J}_d=-n\mu\alpha\nabla n \approx \mu\alpha {n_i}^2/L$.
The characteristic diffusion current, using the Einstein relation for a low density, non-degenerate Bose gas \cite{Ivanov02}, is given by:
$\vec{J}_D=-D\nabla n \approx \mu k_BT n_i/L$.
One expects that the trapping efficiency should fall whenever the size of the external trapping current $\vec{J}_{ext}$ becomes smaller than the internal expanding currents $\vec{J}_{int}=\vec{J}_d+\vec{J}_D$, or in other words when $|\vec{J}_{ext}/\vec{J}_{int}|\le 1$.
As is shown in our results, the behavior of the analyzed system follows this rule-of-thumb fairly well. For the given experimental conditions (with $L/W\approx 3$), the dashed line in Fig \ref{fig:trap_parameter}(e) marks the initial density where $\vec{J}_{ext}\approx\vec{J}_d+\vec{J}_D$, and the dashed line in Fig \ref{fig:trap_parameter}(f) marks the temperature where $\vec{J}_{ext}\approx\vec{J}_d+\vec{J}_D$. There is a substantial decrease in \textit{te} upon increasing density (Fig. \ref{fig:trap_parameter}(e)) and increasing temperature (Fig. \ref{fig:trap_parameter}(f)) above these lines.

These simple and intuitive results can be implemented in the design of devices for trapping and flow control of excitons. For example, one can build a flow device by applying a voltage gradient along a channel \cite{Holleitner06}, and switch the flow on and off by remote dipole interactions with $X_d$ fluids in the control traps, in a device similar to the one depicted in Fig. \ref{fig:device_sketch}(c).

In summary, we propose a method for new dipolar exciton flow devices, which utilizes the remote interactions between spatially separated dipolar exciton fluids. It allows for a direct and model-independent density calibration measurement, as well as for excitonic fluid-fluid control and manipulation. We give simple and intuitive rule-of-thumb estimates for the efficiency of such devices. We also show good agreement between the full numerical model predictions and our simple estimates. Such a concept is quite general and can possibly be implemented in different types of dipolar fluids and other device schemes.

\bibliographystyle{apsrev4-1}
\bibliography{exciton_bibtex}

\end{document}